# The brightest (and faintest) sources in the radio sky

Sarah V. White
Rhodes University
P.O. Box 94, Makhanda 6140, South Africa
sarahwhite.astro@gmail.com

*Abstract*— **Radio observations allow us to identify a wide range of active galactic nuclei (AGN), which are galaxies that have gas accreting onto the supermassive black-hole at the centre. By observing these sources at *multiple* radio frequencies, a more-complete picture can be built of black-hole accretion activity. This completeness is aided by radio waves being unaffected by dust along the line-of-sight to these sources, which cannot be said for waves in the optical part of the electromagnetic spectrum. (Hence, dust obscuration leads to biases in AGN samples selected using optical observations.) A thorough compilation of ~2,000 of the brightest radio-sources in the southern sky [and so of particular relevance for the Square Kilometre Array (SKA) and its precursor/pathfinder telescopes] is the GLEAM 4-Jy (G4Jy) Sample, selected at 151 MHz, with subsets being followed up with MeerKAT and the Australia Telescope Compact Array (ATCA). Meanwhile, optical spectroscopy from the Southern African Large Telescope (SALT) is being used to derive crucial redshift information, both for G4Jy sources and for radio-*faint* AGN in the MeerKAT International GHz Tiered Extragalactic Exploration (MIGHTEE) Survey. The origin of the radio emission in these faint sources is a subject of great debate within the AGN/radio-astronomy communities.**

## I. The role of AGN in Galaxy Evolution

The infall of gas onto a supermassive black-hole typically results in an accretion disc forming, and due to collisions within the disc, a large amount of energy is radiated as *thermal energy* (which in turn leads to gas clouds further out becoming ionised). This is one example of 'AGN feedback' (linking the black hole with its host galaxy), with another being connected to the *kinetic energy* associated with radio jets that are launched from the accretion system. The latter tend to be restricted to 'radio-loud' (i.e. radio-bright) sources, which account for ~10% of the AGN population. We therefore need to study both the radio-loud and 'radio-quiet' (i.e. radio-faint) AGN in order to fully appreciate the influence of feedback processes on how galaxies evolve.

## II. At the radio-bright end: The G4Jy Sample

The G4Jy Sample [1, 2] is over ten times larger than the famous, well-studied 3CRR Sample [3] of the northern hemisphere, and so enables radio-loud AGN properties to be investigated more robustly, and compared with galaxy-evolution/radio-jet simulations, as a function of redshift (i.e. cosmic evolution) and environment. The latter affects the luminosity and morphology of the radio emission, and so needs to be taken into consideration alongside the intrinsic properties of the AGN. For example, the radio-jet axis may be precessing, as is believed to be the case for G4Jy 1523, resulting in a helical structure for the radio emission that is evident through the MeerKAT contours (Fig. 1). The higher spatial resolution afforded by MeerKAT (compared to GLEAM, TGSS, and NVSS) is also crucial for identifying which of the many mid-infrared sources in the vicinity (indicated by green '+' markers in Fig. 1) is the galaxy that hosts the supermassive black-hole giving rise to the extended radio emission. By judging where the radio jets appear to narrow and are centred, we identify the mid-infrared source marked with a white '+' marker in Fig. 1 as the host galaxy for G4Jy 1523. This then allows us to observe the AGN as part of a multi-semester campaign using the Southern African Large Telescope (SALT). This is to obtain optical spectroscopy for the sample (PI: White), from which we can calculate redshifts (i.e. how distant the galaxies are) and intrinsic properties for this legacy dataset, and in doing so make these powerful radio-sources 'science-ready' for the SKA.

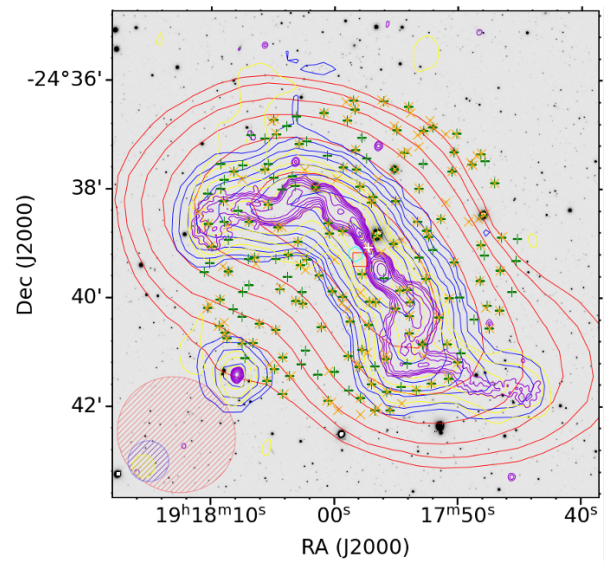

Fig. 1. Radio contours indicating the spatial distribution on the sky (RA = Right Ascension, Dec = Declination, in J2000 celestial coordinates), and intensity, of the radio emission for G4Jy 1523. Contours derived from GLEAM (red), TGSS (yellow), and NVSS (blue) data are overlaid onto a mid-infrared image from AllWISE (inverted greyscale). The host galaxy is marked with a white '+', and the excellent sensitivity and resolution of MeerKAT is demonstrated through the purple contours, obtained with < 5 min of integration time per G4Jy source [4]. Beam sizes are shown in the bottom-left corner.

The G4Jy Sample was selected at 151 MHz using MWA observations, with a total of 20 flux-density measurements available across 72–231 MHz. Further to MeerKAT (1.3 GHz) follow-up, we have ATCA (5, 9, 20 GHz) observations that enable us to construct broadband radio spectra. The straight lines in Fig. 2 illustrate the degree of error that may arise if a power-law function ($S_\nu \sim \nu^\alpha$) is used to extrapolate/interpolate the flux density, $S_\nu$ (in units of Jy, where 1 Jy = $10^{-26}$ W m$^{-2}$ Hz$^{-1}$), at different frequencies, $\nu$, with spectral index, $\alpha$. In this case, the spectral shape is described as a 'peaked spectrum', which is typical of radio galaxies in the early phase of their lifetime.

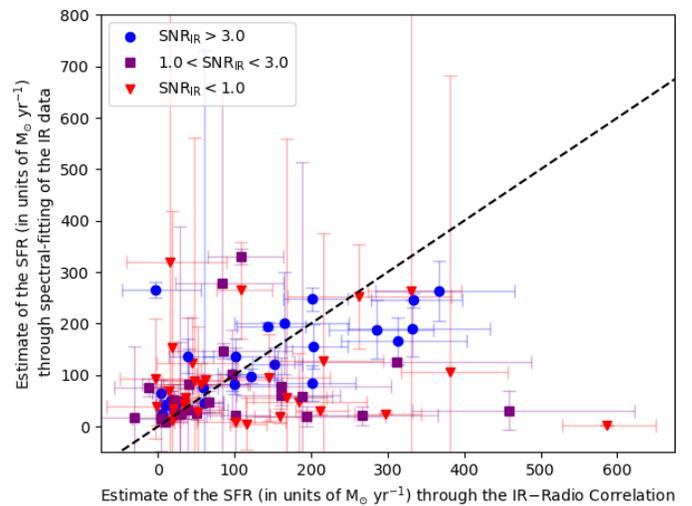

Fig. 3. A comparison of SFRs for quasars in the MIGHTEE Survey [5, 6], colour-coded by the signal-to-noise ratio of the IR data, SNR$_{IR}$ [10]. The dashed line is to guide the eye where the two SFR estimates are equal to one another.

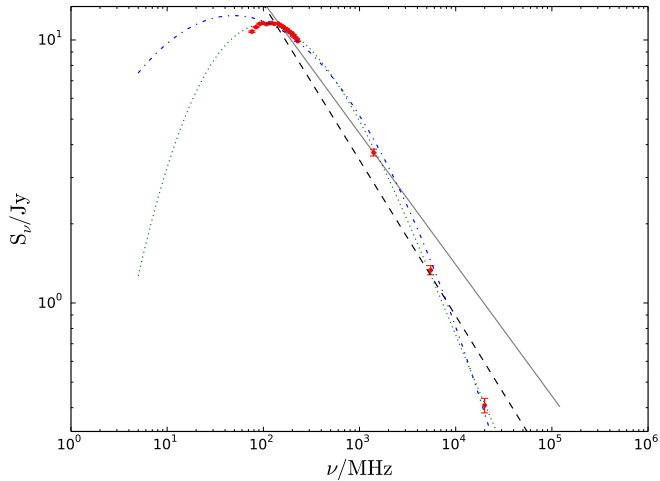

Fig. 2. A log-log plot showing an example of a broadband radio spectrum with flux-density measurements (from 72 MHz to 20 GHz; red datapoints) from the Murchison Widefield Array (MWA), the NVSS, and ATCA. The horizontal-axis range spans 1 MHz to 1 THz, and the curved lines (dash-dotted and dotted) are 2$^{\text{nd}}$- and 3$^{\text{rd}}$-order polynomial fits to the data, respectively.

### III. AT THE RADIO-FAINT END: MIGHTEE QUASARS

Over 90% of the AGN population have relatively low levels of radio emission, with an absence of the powerful radio-jets that are ubiquitous in, e.g., the G4Jy Sample. At these faint flux-densities it can be difficult to disentangle the star-formation contribution to the total radio emission from the black-hole accretion contribution, but the incorporation of infrared (IR) data in the analysis (Fig. 3) can help to break this degeneracy. This is because IR emission from a galaxy provides an independent route for measuring the star-formation rate (SFR), which is expressed in units of solar masses per year, $M_\odot$ yr$^{-1}$. We compare this value (along the ordinate axis of Fig. 3) with an SFR estimate that *assumes* that all of the radio emission (detected via MIGHTEE [5, 6]) can be attributed to star formation, and so lies on the IR–radio correlation determined by [7] for star-forming galaxies (abscissa axis of Fig. 3). If this assumption is incorrect then the datapoint lies below the dashed line in Fig. 3, indicating that the quasar must in fact have an additional contribution towards the radio emission, thus providing evidence for AGN-related radio emission [8, 9, 10].

ACKNOWLEDGEMENT

SVW acknowledges the financial assistance of the South African Radio Astronomy Observatory (SARAO; https://www.sarao.ac.za), and thanks the anonymous reviewers, whose comments helped to improve the paper.